\documentclass[%
reprint,
superscriptaddress,
 amsmath,amssymb,
 aps,
pra,
]{revtex4-2}

\usepackage{graphicx}
\usepackage{dcolumn}
\usepackage{bm}
\usepackage{hyperref}
\hypersetup{
    colorlinks=true,
    linkcolor=black,
    citecolor=blue,
    urlcolor=black
    }

\DeclareMathOperator{\erf}{erf}
\DeclareMathOperator{\erfc}{erfc}

\begin{document}

\preprint{APS/123-QED}

\title{Mitigating Scattering in a Quantum System Using Only an Integrating Sphere}

\author{Zhenfei Jiang}
\affiliation{Institute for Quantum Science and Engineering, Texas A$\&$M University, College Station, TX 77843, USA}
\affiliation{Department of Physics and Astronomy, Texas A$\&$M University, College Station, TX 77843, USA}%

\author{Tian Li}
\email{tian-li@utc.edu}
\affiliation{Department of Chemistry and Physics, University of Tennessee at Chattanooga, Chattanooga, TN 37403, USA}
\affiliation{UTC Research Institute, University of Tennessee at Chattanooga, Chattanooga, TN 37403, USA}

\author{Matthew L. Boone}
\affiliation{Department of Chemistry and Physics, University of Tennessee at Chattanooga, Chattanooga, TN 37403, USA}

\author{Zhenhuan Yi}
\email{yzh@tamu.edu}
\affiliation{Institute for Quantum Science and Engineering, Texas A$\&$M University, College Station, TX 77843, USA}
\affiliation{Department of Physics and Astronomy, Texas A$\&$M University, College Station, TX 77843, USA}%

\author{Alexei V. Sokolov}
\affiliation{Institute for Quantum Science and Engineering, Texas A$\&$M University, College Station, TX 77843, USA
}%
\affiliation{Department of Physics and Astronomy, Texas A$\&$M University, College Station, TX 77843, USA
}%
\affiliation{Department of Physics, Baylor University, Waco, TX 76798, USA 
}%

\author{Girish S. Agarwal}
\affiliation{Institute for Quantum Science and Engineering, Texas A$\&$M University, College Station, TX 77843, USA
}%
\affiliation{Department of Physics and Astronomy, Texas A$\&$M University, College Station, TX 77843, USA
}%
\affiliation{Department of Biological and Agricultural Engineering, Texas A$\&$M University, College Station, TX 77843, USA
}%

\author{Marlan O. Scully}
\affiliation{Institute for Quantum Science and Engineering, Texas A$\&$M University, College Station, TX 77843, USA
}%
\affiliation{Department of Physics and Astronomy, Texas A$\&$M University, College Station, TX 77843, USA
}%
\affiliation{Department of Electrical and Computer Engineering, Princeton University, Princeton, NJ 08540, USA
}%

\date{\today}

\begin{abstract}
Strong quantum-correlated sources are essential but delicate resources for quantum information science and engineering protocols. Decoherence and loss are the two main disruptive processes that lead to the loss of nonclassical behavior in quantum correlations.
In quantum systems, scattering can contribute to both decoherence and loss. In this work, we present an experimental scheme capable of significantly mitigating the adverse impact of scattering in quantum systems. Our quantum system is composed of a two-mode squeezed light generated with the four-wave mixing process in hot rubidium vapor, and a scatterer is introduced to one of the two modes. An integrating sphere is then placed after the scatterer to recollect the scattered photons. We use mutual information between the two modes as the  measure of quantum correlations, and demonstrate a 47.5~\% mutual information recovery from scattering, despite an enormous photon loss of greater than 85~\%. Our scheme is the very first step towards recovering quantum correlations from disruptive random processes, thus has the potential to bridge the gap between proof-of-principle demonstrations and practical real-world implementations of quantum protocols. 
\end{abstract}

\maketitle


\section{\label{sec:level1}Introduction}

Quantum information science and engineering (QISE) harness the principles of quantum mechanics to manipulate and process information, leading to groundbreaking advancements in computing, communication, and sensing~\cite{nielsen2010quantum}. Nevertheless, quantum correlations, the cornerstone of QISE, are inherently fragile. They are very susceptible to disruptive processes such as decoherence and loss, resulting in the loss of their non-classical behaviors~\cite{joos2013decoherence}. Therefore, mitigating the adverse effects of those disruptive processes on quantum correlations in quantum systems is highly beneficial for the implementations of QISE protocols~\cite{bachor2019guide}. 

Among those well studied quantum systems, squeezed states of light have been used as reliable quantum resources
in various QISE protocols~\cite{li2022quantum,casacio2021quantum,de2020quantum,lawrie2019quantum,michael2019squeezing,clark2016observation,aasi2013enhanced,taylor2013biological},
particularly those devised in the continuous variable (CV) regime~\cite{braunstein2005quantum}. 
 Here, we use a two-mode squeezed state of light generated with the four-wave mixing (FWM) process in hot rubidium vapor as the quantum source to study mitigating the adverse effects of disruptive processes on quantum correlations. It is well understood that squeezing level is equally reduced by linear loss, while optical group delay can be corrected electronically without affecting the squeezing, e.g., the delay between the probe and conjugate mode of our source due to refractive index difference. It is also worth noting that active shaping of twin photons' temporal wave package can revive entanglement by restoring quantum interference \cite{Wu2019}. However, real-world applications, such as quantum sensing and imaging, often encounter absorption and random scattering. To simplify the model, we employ a scatterer to introduce disturbance to the quantum system because scattering can contribute to both decoherence and scattering loss. 
 Mitigation is realized by simply recollecting scattered photons using an integrating sphere (IS).

We adopt mutual information (MI) as the information metric to measure quantum correlations~\cite{serafini2003symplectic}. MI and its quantum counterpart, quantum mutual information (QMI), are very effective tools in quantifying quantum correlations~\cite{clark2014quantum,vogl2014advanced,weedbrook2012gaussian}. This metric provides a precise measure of the shared MI between two quantum systems, offering valuable insights into their quantum correlations. In fact, QMI in bipartite Gaussian states in the CV regime have been employed to facilitate a coherent representation of information within quantum systems, thus fostering valuable applications in quantum information processing~\cite{ameri2015mutual,dixon2012quantum,wolf2008area,belavkin2002entanglement}. 
\textcolor{black}{Since calculating the QMI for bipartite Gaussian states involves determining the covariance matrix and this work does not involve field quadrature measurements, we therefore did not use the QMI as the measure of quantum correlations. Instead, we use Shannon entropy to calculate the MI~\cite{serafini2003symplectic} for our quantum system consisting of the quantum-correlated bright two modes generated through the FWM process.} Specifically, \textit{the MI is calculated from the joint probability of the two modes' intensity fluctuations}, which encapsulate the squeezed nature between them. A ground glass diffuser, acting as a photon scatterer, is  mounted at the front aperture of an IS, which then collects the scattered photons and sends them through its back aperture to a photo-detector. Using this scattering mitigation scheme, we are able to recover $47.5~\%$ MI after a time delay of 32.7~ns, despite an enormous photon loss of greater than $85~\%$. Due to the simplicity of our scheme, which only involves an IS, it can find many applications in QISE protocols where scattering is inevitable. 


\section{Experiment}

\begin{figure}
\centering
\includegraphics[width=\linewidth]{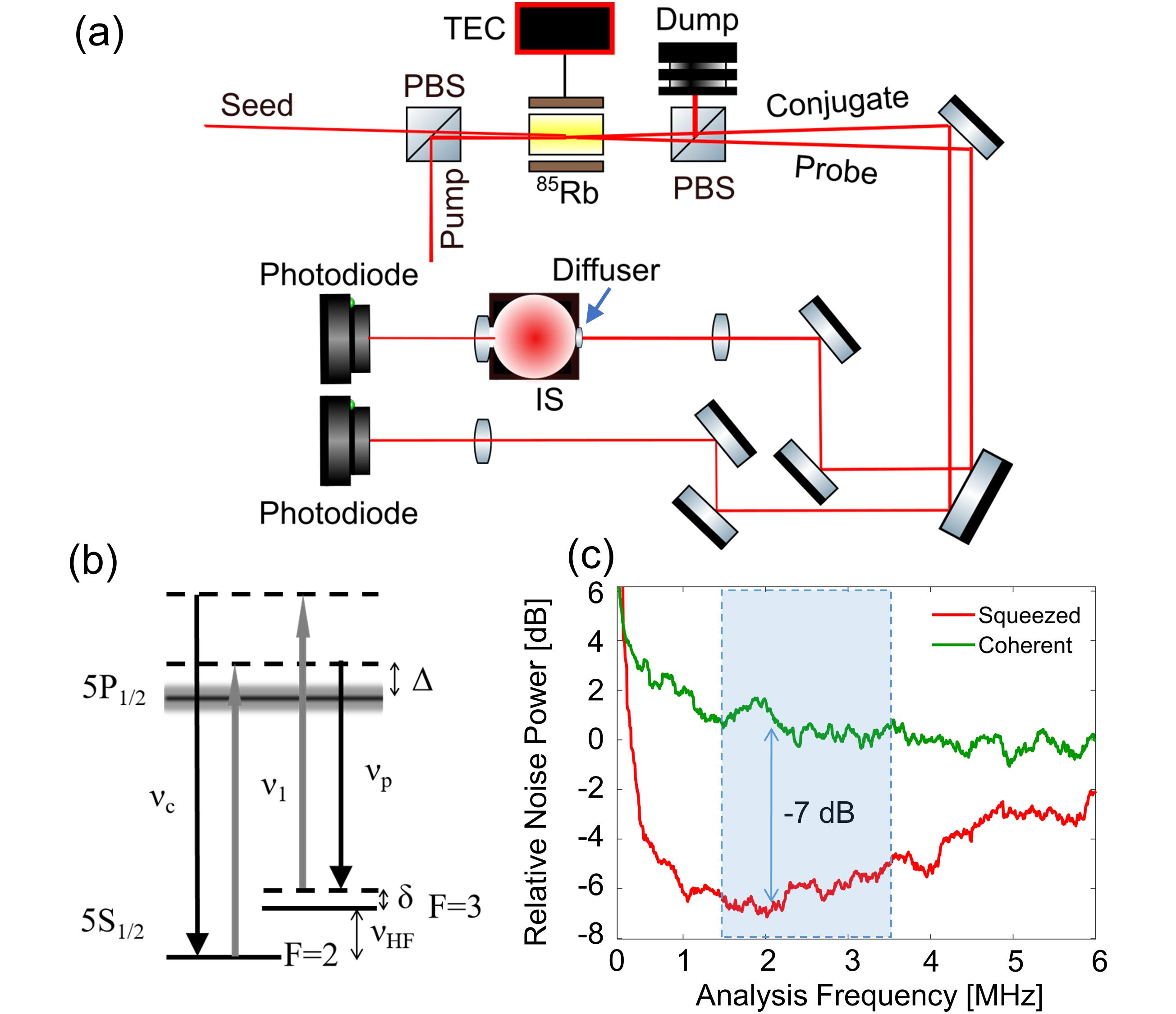}
\caption{(a) Experimental setup in which a seeded $^{85}$Rb vapor cell generates strong quantum-correlated `twin' beams, i.e., the `probe' and `conjugate' beams, via FWM. A 5-cm-diameter integrating sphere (IS) is placed on the probe beam path, and a `scatterer', i.e., a half-inch-diameter ground glass diffuser, is placed at the entrance aperture of the IS. The focused probe beam is incident on the specular side of the diffuser. The other side (facing the IS) of the diffuser has 120 grits which can produce near-spherical scattering patterns. A lens is attached to the exit aperture of the IS to collect escaped probe photons from IS and send them to a homemade photodetector. PBS: polarizing beam splitter, TEC: thermoelectric cooler. (b) Energy diagram of the $\text{D}_1$ transition of $^{85}$Rb atom. The optical transitions form a double-$\Lambda$ configuration, where ${\nu}_1$, ${\nu}_p$, and ${\nu}_c$ denotes the pump, probe, and conjugate frequencies, respectively, with ${\nu}_p+{\nu}_c =2{\nu}_1$. The Doppler broadening is represented by the width of the excited state (gray). $\Delta$ and $\delta$ are the single- and two-photon detuning, respectively. ${\nu}_{HF}$ is the hyperfine splitting in the 5S$_{1/2}$ state of $^{85}$Rb. (c) A typical two-mode intensity-difference squeezing spectrum obtained by post-processing the intensity fluctuations of the twin beams in software. The red and green curves denote the spectra of the squeezed and coherent beams, respectively. Squeezing level of 7 dB can be achieved on daily basis. The shaded area from 1.5~MHz to 3.5~MHz is our region of interest for data analysis.
\label{fig:setup}}
\end{figure}

The two-mode squeezed light used in this work is generated with the FWM process in an atomic $^{85}$Rb vapor cell~\cite{li2022quantum,dowran2018quantum,anderson2017phase,pooser2015ultrasensitive,clark2014quantum,hudelist2014quantum}. This squeezed light generation scheme has proven to be an effective way of producing quantum correlations~\cite{li2021experimental,li2017improved,clark2014quantum,vogl2014advanced,vogl2013experimental}. The experimental setup and the respective $^{85}$Rb atomic level structure are shown in Fig.~\ref{fig:setup}(a) and (b). The atomic medium is pumped by a strong ($\sim 500$~mW) narrow-band continuous-wave (CW) laser at frequency $\nu_1$ ($\lambda = 795$~nm) with a typical line-width $\Delta \nu_1 \sim 100$~kHz. Applying an additional weak (from a few hundred $\mu$W to $\sim 1$~mW) coherent seed beam 
at frequency $\nu_p = \nu_1 - (\nu_{HF}+\delta)$, where $\nu_{HF}$ and $\delta$ are the hyperfine splitting in the electronic ground state of $^{85}$Rb and the two-photon detuning, respectively, in Fig.~\ref{fig:setup}(b) (further experimental details can be found in Ref.~\cite{li2022quantum}). Two pump photons are converted into a pair of `twin' photons, namely `probe $\nu_p$' and `conjugate $\nu_c$' photons, adhering to the energy conservation $2 \nu_1 = \nu_p + \nu_c$ (see the level structure in Fig.~\ref{fig:setup}(b)). The photon number fluctuations (i.e., intensity fluctuations) of the probe and conjugate beams are strongly quantum-correlated exhibited by a strong two-mode squeezing between them. 

A typical squeezing spectrum is shown in Fig.~\ref{fig:setup}(c) obtained by post-processing the intensity fluctuations of the probe and conjugate beams in software. \textcolor{black}{The shot noise limit (SNL), shown as the green curve in Fig.~\ref{fig:setup}(c), was measured by isolating the `seed' beam (a coherent beam, labeled as `Seed' in Fig.~\ref{fig:setup}(a)) before it entered the $^{85}$Rb cell. We split the beam using a half-wave plate and a polarizing beam splitter (PBS), ensuring that the two outputs of the PBS matched the optical powers of the twin beams. These two resulting coherent beams were then directed into two photo-detectors to register their intensity fluctuations. These detectors are home-built, and their photo-diodes have quantum efficiency greater than 92~\%. The intensity fluctuation time traces were subtracted, and then Fourier transformed in software to yield the noise power of the intensity difference between the two beams. This noise power level serves as a measure of the SNL for the total optical power arriving at the two photo-detectors. The SNL is expected to be independent of frequency, which is consistent with our observations within the bandwidth of the detection electronics, with a drop-off occurring above 10~MHz (not shown in Fig.~\ref{fig:setup}(c)). The red curve in Fig.~\ref{fig:setup}(c) represents the noise power for the intensity difference between the probe and conjugate beams, showing a 7~dB reduction below the SNL.}

\textcolor{black}{Note that in this work, we are not measuring squeezing in field quadratures, but rather in the intensity difference of a bipartite entangled state, which is generated through the FWM process in $^{85}$Rb by converting two photons from a strong coherent pump beam into twin photons emitted into spatially separated probe and conjugate mode. This type of squeezing, also referred to as ``\textit{bright two-mode squeezing},'' results in a photon flux of the probe and conjugate beams ranging from $10^{13}$ to $10^{16}$ photons per second~\cite{li2020squeezed}. Although the generation of photons in each mode is random (i.e., thermal-like individually), there is a strong quantum correlation between the probe and conjugate modes in their intensity fluctuations because the entangled photons are produced in pairs by the FWM process. This strong quantum correlation leads to the observed squeezing in the intensity difference of the probe and conjugate beams shown in Fig.~\ref{fig:setup}(c).}

Since the pump and the twin beams are cross-polarized, thus after the cell, we use a Glan-Laser polarizing beam splitter with $\sim 2\times 10^5:1$ extinction ratio for the pump polarization, to filter out the pump photons from the twin beams. The conjugate beam is then sent directly to the home-built photo-detector, while the probe beam propagates through a scatterer, i.e., a half-inch-diameter ground glass diffuser (Thorlabs, DG05-120) with its polished side facing the beam. The other side of the scatterer is a medium-grit surface (120 grit), producing near-spherical scattering patterns. The scatterer is placed at the entrance aperture (also half-inch-diameter) of a 5-cm-diameter IS (Thorlabs, IS200-4), with which almost all scattered photons can be confined inside. After the IS, a one-inch-diameter lens with 25~mm focal length is attached immediately to the exit aperture of the IS to collect those exiting photons and focus them to another lab-made photo-detector with a quantum efficiency similar to the conjugate photo-diode. Due to repeated imperfect reflections of the scattered photons inside the IS with $\sim$ 98~$\%$ surface reflectivity, optical power of the exiting photons is only $\sim$ 14 $\%$ of the probe input power.



We used an oscilloscope to record photo-currents from the photo-detectors for post-processing the quantum correlations in the photon number fluctuations (i.e., intensity fluctuations) of the twin beams.  \textcolor{black}{Also note that since we are focused on the temporal quantum correlations in the total intensities of the twin beams, the polarization of the twin beams is not relevant to our study}. The registered photo-current (i.e., oscilloscope trace of voltage vs time) of each detector consists of 4 million points acquired at a sampling rate of 2.0 giga-samples per second for a total acquisition time of 2~ms. In the following, we use the MI formalism to describe the correlations of the twin beams and provide a theoretical model that explains our experimental observations quite well.\par


\section{Results}

\subsection{Mutual Information of Twin Beams}

\begin{figure*}
\centering
\includegraphics[width=\linewidth]{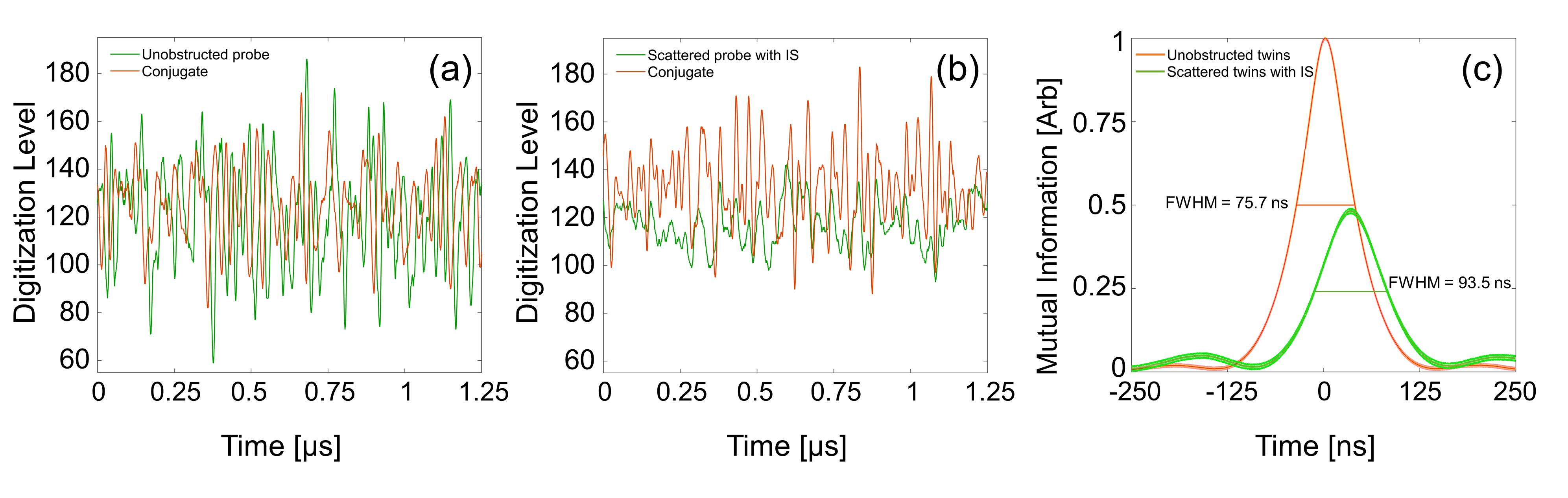}
\caption{Intensity fluctuations of the twin beams in (a) and (b) are for experimental configurations described in the text as (i) and (ii), respectively. (c) Normalized MI of the twin beams as functions of relative delay obtained by calculating Eq.~(\ref{eq1}) at every time-shift step of 0.5~ns. The red and green curves are the MI for configuration (i) and (ii), respectively. The shadow on each curve represents one standard deviation of averaging over 10 pairs of intensity fluctuations. 
\label{fig:mi1} }
\end{figure*}

The MI of the twin beams is calculated under two configurations: (i) probe and conjugate beams are unobstructed, and (ii) a combination of the scatterer and the IS is placed on the probe beam path, while the conjugate beam is unobstructed. Optical power of the probe beam before the scatterer is 5.9 mW, which reduces to 750~$\mu$W after passing the scatterer and being re-collected by the IS, consisting only 14~$\%$ of the input probe power. In both configurations, optical power of the conjugate beam is kept at 5.3~mW. The intensity fluctuations of the twin beams in configurations (i) and (ii) are shown in Fig.~\ref{fig:mi1}(a) and Fig.~\ref{fig:mi1}(b), respectively, where x-axis is time, and y-axis is the digitization level of an 8-bit oscilloscope with maximal value of $2^8=256$.

As can be seen in Fig.~\ref{fig:mi1}(a), when the twin beams are unobstructed, i.e., without the scatterer on the probe beam path, the temporal behaviors of the  twin beams' intensities are quite similar, even among very fine (i.e., high frequency) fluctuating features. This indicates the existence of strong correlations between the unobstructed probe and conjugate beams beyond classical noise regime. In Fig.~\ref{fig:setup}(c), the quantum correlations in the shot noise limited regime are manifested by more than 7~dB intensity-difference squeezing.  For configuration (ii), the level range of the probe intensity fluctuation shown in Fig.~\ref{fig:mi1}(b) is significantly reduced along with the probe optical power due to the proportionality between the variance and the mean. The level range of intensity fluctuations of the conjugate beam stays the same. As can be seen in Fig.~\ref{fig:mi1}(b), the randomness introduced to the probe intensity fluctuations by the scatterer visibly breaks the intensity correlations in large time scales, i.e., classical correlations in the technical noise regime. It also washes out the correlated fine fluctuating features, i.e., quantum correlations in the shot noise regime, making the temporal behavior of the probe beam intensity appear to be smoother after traveling through the scatterer and the IS.


To compute the MI, the intensity fluctuations of the probe and conjugate beams are filtered in software by a band-pass filter from 1.5~MHz to 3.5~MHz, where the source has the best squeezing level, as shown in Fig.~\ref{fig:setup}(c) by the shaded area on the squeezing spectrum. Despite individually, the probe and conjugate fluctuations are random in nature,  fluctuations on one beam carry information about the fluctuations on the other. Therefore, correlated information is shared between the twin beams, even to the shot noise level.
We capture these correlations by calculating the MI between the intensity fluctuations of the twin beams using the following equation~\cite{nielsen2010quantum}: 

\begin{equation}
    I(p;c) =  \sum_{1}^{N_p} \sum_{1}^{N_c} P(p,c) \log_2 \frac{P(p,c)}{P(p)P(c)},
    \label{eq1}
\end{equation}
where $P(p,c)$ is the joint probability distribution obtained from a 2-D intensity fluctuation histogram by binning the intensity fluctuations of the probe and conjugate beams individually, i.e., \textit{binning the intensity fluctuations on the y-axes} in Figs.~\ref{fig:mi1}(a) and (b). $P(p)$ and $P(c)$ are the marginal probabilities of the intensity fluctuations of the probe and conjugate beams, respectively. $N_p$ and $N_c$ are the number of bins for the two intensity fluctuations. We then can obtain the MI as a function of time by shifting the two intensity fluctuations relative to each other along the x-axis by a step of 0.5~ns (i.e., the sampling resolution of the oscilloscope), and calculate Eq.~(\ref{eq1}) at each time shift. Note that, in contrast to the calculation of QMI which would involve the knowledge of both quadratures of the two fields, Eq.~(\ref{eq1}) only yields the MI in the amplitude quadrature. 


The number of bins can vary depending on the dynamic range of the intensity fluctuations, i.e., the number of digitization levels the fluctuations occupy. With a finite dynamic range, the larger the number of bins, the finer the bin size will be, we thus are able to resolve fine/high frequency fluctuations. As the dynamic range of most of the intensity fluctuations are within 100 digitization levels, we thus pick $N_p=N_c=100$ throughout our data analysis for obtaining histograms. 
\par

The MI of the twin beams as functions of time shift (i.e, Eq.~(\ref{eq1}) is calculated at each time-shift step) for cases (i) and (ii) are presented in Fig.~\ref{fig:mi1}(c) by the red and green curves, respectively. Each curve is the average over 10 pairs of intensity fluctuations. Note that error bars representing one standard deviation are also shown on the graph, although they are too small to be readily discerned. We normalize the peak height of the MI in case (i) to unity; it appears to look like a Gaussian profile with a full width at half maximal (FWHM) of $75.7$~ns. With this normalization, in case (ii), the peak height of the MI is reduced to 0.475 with a broadened FWHM of $93.5$~ns. 
In addition to the Gaussian profile change, the peak position of the green curve is also delayed by $\tau_0 = 32.7$~ns with respect to the peak position of the red curve. This time delay $\tau_0$ can be comprehended as the ``memory time'' of the IS, implying that 47.5~\% of the input (i.e., unobstructed) MI can be stored in the IS for $\tau_0 =32.7$~ns.

We also investigated the configuration in which \textit{only the scatterer is present} on the probe beam path. We used the same experimental setup shown in Fig.~\ref{fig:setup}(a) but \textit{without the IS}. The optical power is insufficient to produce meaningful intensity fluctuations, as these are overwhelmed by the detector’s electronic noise. Based on this experimental observation, it is reasonable to assume that without the IS, the MI would be largely lost due to scattering.

\textcolor{black}{An intuitive explanation of how an IS mitigates scattering can be provided as follows. Because of the diffuser, the probe photons are scattered nearly uniformly in all directions, covering a solid angle of $4\pi$ steradians. By placing an IS \textit{immediately} after the diffuser, the forward-scattered probe photons from all directions are effectively recollected by the IS. These photons then undergo multiple reflections on the inner spherical surface of the IS before escaping through the exit port and eventually being focused on the photo-diode. Due to these multiple reflections, a time delay ($\tau_0 = 32.7$~ns) in the MI peak relative to the unobstructed MI is observed. Additionally, due to the inner surface of the IS having a reflectivity of approximately 98~\%, some probe photons are inevitably lost. This results in a degraded MI peak, which is 47.5~\% of the unobstructed MI peak.}

\begin{figure*}
\centering
\includegraphics[width=\linewidth]{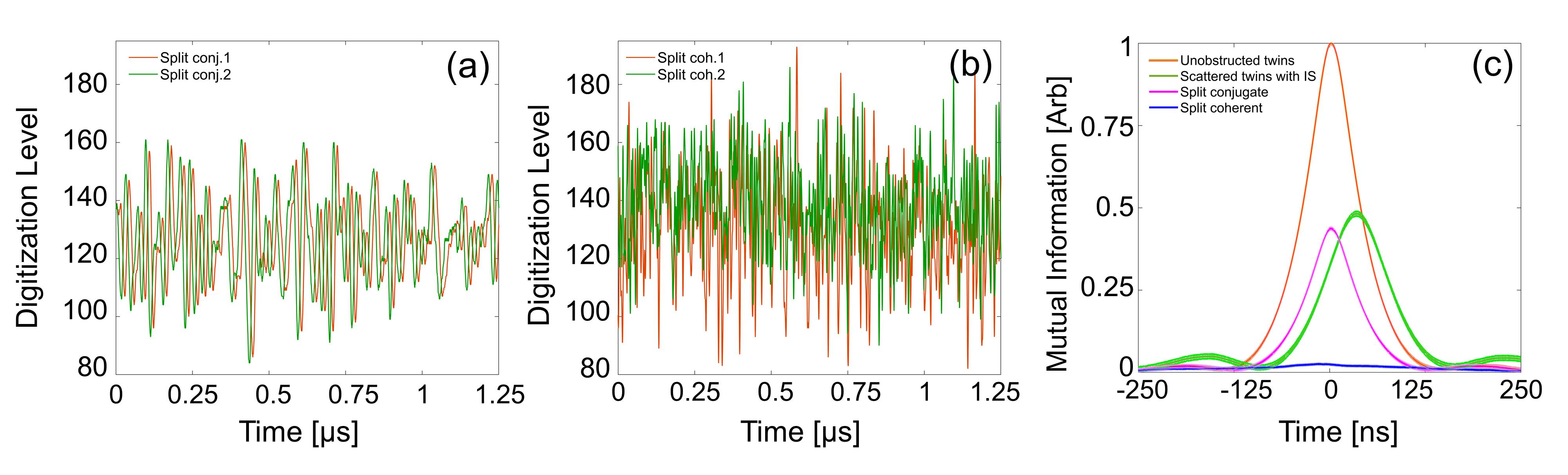}
\caption{(a) Intensity fluctuations of split conjugate beams. (b) Intensity fluctuations of split coherent beams. The magenta and blue curves in (c) are the MI of split conjugate beams and split coherent beams, respectively. They are plotted on top of Fig.~\ref{fig:mi1}(c) for comparison. The shadow on each curve represents one standard deviation of averaging over 10 pairs of intensity fluctuations. 
\label{fig:mi2}}
\end{figure*}

\subsection{Mutual Information of Split Conjugate Beams and Split Coherent Beams}

\textcolor{black}{Since the probe beam contains both the coherent `seed' photons and the thermal-like probe photons generated by the FWM process~\cite{li2021experimental}. Consequently, the probe beam, as well as the conjugate beam, exhibit a mixed statistical nature with convolved coherent light (Poisson) and thermal light (Bose-Einstein) distributions. Therefore, the second-order correlation function $g^{(2)}$ for the probe and conjugate beams would lie between 1 and 2, where $g^{(2)}=1$ corresponds to coherent light and $g^{(2)}=2$ corresponds to thermal light. The measured $g^{(2)}$ values for our probe and conjugate beams are $1.146 \pm 0.012$ and $1.238 \pm 0.011$, respectively. Since the conjugate beam exhibits the most thermal-like behavior, we therefore use the ``split conjugate'' beams for comparison to demonstrate the amount of MI residing in two \textit{classically} correlated beams.}


We used a $\lambda/2$-waveplate in conjunction with a polarizing beamspliter (PBS) to split the conjugate beam so that the two outputs of the PBS have the same optical power as the unobstructed twin beams. Neither the scatterer or the IS is present on the beam paths. We plot the intensity fluctuations of the two split conjugate beams in Fig.~\ref{fig:mi2}(a). The MI of the split conjugate beams is plotted in Fig.~\ref{fig:mi2}(c) as the magenta curve. As can be seen from Fig.~\ref{fig:mi2}(a) that although the two split conjugate beams clearly exhibits correlated fluctuations, they are visibly \textit{smoother} than the fluctuations of the unobstructed twin beams shown in Fig.~\ref{fig:mi1}(a), indicating a lack of high frequency (i.e., shot-noise limited) correlations between the split conjugate beams, which is manifested by the significantly lowered peak height of the magenta curve relative to the red curve in Fig.~\ref{fig:mi2}(c). Also note that since there is no IS present in the split conjugate beams, the peak of the magenta curve lines up quite well with the peak of the red curve. \textcolor{black}{The additional MI peak height of the twin beams compared to the split conjugate beams indicates the presence of quantum correlations in the twin beams \textit{beyond the classical correlations} in the split conjugate beams.}

\textcolor{black}{We also applied identical procedures to measure the MI between two split coherent beams. The intensity fluctuations of the split coherent beams are shown in Fig.~\ref{fig:mi2}(b), since there are no appreciable correlations exhibited in the shot-noise limited frequency regime, it hence gives rise to a close to zero MI, represented by the blue curve in Fig.~\ref{fig:mi2}(c).}


\textcolor{black}{Note that in this work, the temporal quantum correlations are generated in the continuous-wave (CW) regime from a seeded FWM process in $^{85}$Rb atomic vapor. These correlations are centered on the $^{85}$Rb $\text{D}_1$ line and exhibit a spectral width of approximately 10~MHz~\cite{li2022quantum,clark2014quantum}. However, as noted in Ref.~\cite{fry2006integrating}, when using a pulsed laser, the repetition rate must be slow enough to allow the duty cycle to accommodate the exponential decay of the input pulse, which is influenced by the geometry of the IS. For both CW and pulsed lasers, the bandwidth consideration is crucially linked to the reflectivity of the IS’s inner surface. Ensuring that the input light's bandwidth matches the spectral response of the IS’s high reflectivity is a key condition for our scattering mitigation scheme to work effectively. We therefore attribute the main limitation of our scheme to the imperfect reflectivity of the inner surface of the IS. While the single pass reflectivity is 98~\%, multiple reflections from the surface induce significant overall photon loss. We anticipate that using an IS with near-unity inner surface reflectivity would significantly improve our current MI recovery performance from 47.5~\% to a substantially higher value.\\}

\section{Theoretical Analysis}

\begin{figure}
\centering
\includegraphics[width=\linewidth]{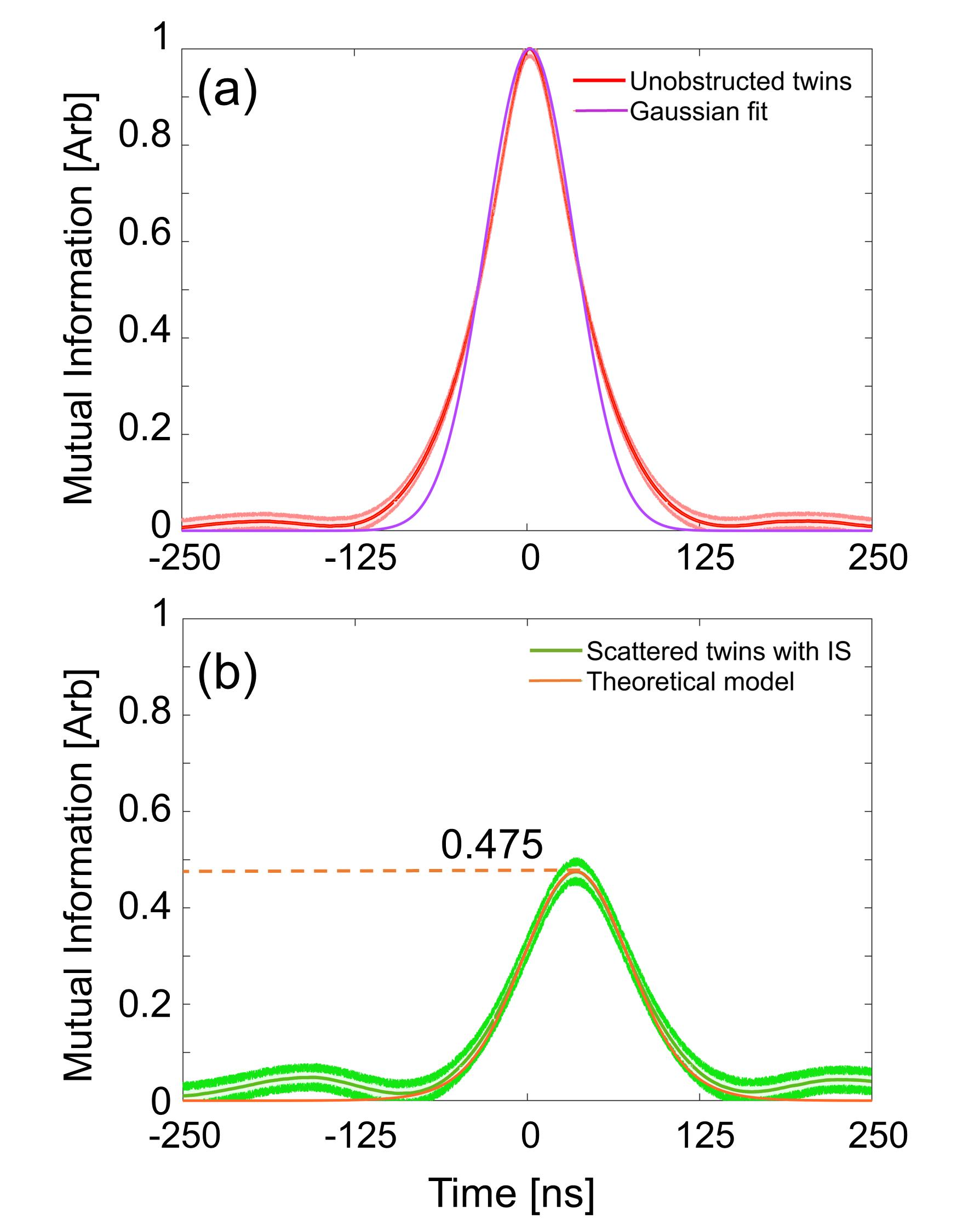}
\caption{The Gaussian fit described by Eq.~(\ref{eq:2}) is shown in (a) as the purple curve. The theoretical model described by Eq.~(\ref{eq:5}) is shown in (b) as the orange curve. The red and green curves are the same experimental data shown in Fig.~\ref{fig:mi1}(c). Both the Gaussian fit and the theoretical model are in good agreement with the experimental data.}
\label{fig:theory}
\end{figure}

Our theoretical analysis for the experimental observations shown in Fig.~\ref{fig:mi1}(c) based upon the assumption that the unobstructed two-mode squeezed state is Gaussian, i.e., it can be described by a Gaussian Wigner function~\cite{weedbrook2012gaussian,serafini2003symplectic,duan2000inseparability}. We thus fit the MI of the unobstructed twin beams, i.e., the red curve shown in Fig.~\ref{fig:mi1}(c), using a Gaussian function\cite{duan2000inseparability,vogl2013experimental,weedbrook2012gaussian}:
\begin{equation}
        {g(t)}
        =  {\exp \left( -\dfrac{t^2}{2{\sigma_0}^2} \right)},
        \label{eq:2}
\end{equation}
where $\sigma_0$ determines the theoretical value of the red curve's FWHM, which is given by 2$\sqrt{2\ln2}\sigma_0$. The Gaussian fit expressed by Eq.~(\ref{eq:2}) is shown in Fig.~\ref{fig:theory}(a) as the purple curve, from which, we obtain $\sigma_0$ = 32.1~ns.

After passing through the diffuser, scattered probe photons in all directions go through Lambertian diffusive reflections on the IS's inner surface many times before a fraction eventually escape from the IS through its apertures. These repeated random reflections give rise to a random delay in the arrival of the scattered probe photons at the exit aperture of the IS. \textcolor{black}{A detailed theoretical analysis of the IS is presented in Ref.~\cite{fry2006integrating}, which includes calculations of the IS's temporal response and experimental results on the transformation of a pulse by the IS. The study indicates that the reflectivity of the inner surface of the IS is nearly unity (98~\%) and that the output pulse can be described by convolving the input with an \textit{exponential} distribution. Our theoretical modeling of the MI data is based on these key characteristics of the IS. The parameters of our theoretical model are determined by the design of the IS, and we extract these parameters from our data. We hence write the MI of the output probe and conjugate beams as} 
\begin{equation}
    {G(t)}
    = 
    \eta {\int_{-\infty}^{+\infty} g(t- \tau) p(\tau) \,d\tau},
    \label{eq:3}
\end{equation}
where $\eta$ is the forward scattering efficiency \textcolor{black}{of the diffuser} (as opposed to backward scattering), and $p(\tau)$ is the characteristic temporal function for the system composed of a scatterer and an IS, featuring an exponential distribution according to Ref.~\cite{fry2006integrating}:
\begin{equation}
    {p(\tau)}
    =
    {\dfrac{1}{2\sigma} \exp{\left[-\dfrac{ |\tau -\tau_0|}{\sigma} \right]}}.
    \label{eq:4}
\end{equation}
Here $\tau_0$ and $\sigma$ are the average and uncertainty of the temporal delay, respectively. We have three unknown parameters $\eta$, $\tau_0$, and $\sigma$ which need to be fixed from measurement data. We first note that using Eq.~(\ref{eq:2}) and Eq.~(\ref{eq:4}), the function G in Eq.~(\ref{eq:3}) can be analytically evaluated in terms of the error function $\erf(x)=(2/\sqrt{{\pi}}){\int_{0}^{x} \exp(-t^2) \,dt}$:
\begin{equation}
\begin{aligned}
     G(T) &= \frac{\eta \sigma_0\sqrt{2\pi}}{4\sigma} \exp \left( \frac{\sigma_0^2}{2\sigma^2} \right)\\
     & [\exp \left( -\frac{T}{\sigma} \right) \erfc \left( \frac{\sigma_0^2 - T\sigma}{\sqrt{2}\sigma\sigma_0} \right) \\
    & + \exp \left(\frac{T}{\sigma} \right) \erfc \left( \frac{\sigma_0^2 + T\sigma}{\sqrt{2}\sigma\sigma_0} \right) ],
\end{aligned}
\label{eq:5}
\end{equation}
where $T=t-\tau_0$ and $\erfc(x)=1-\erf(x)$. The parameter $\tau_0$ is obtained from the peak position shift in Fig.~\ref{fig:mi1}(c) with the value 32.7 ns. The peak value is give by 
\begin{equation}
    G(t=\tau_0) = \frac{\eta \sigma_0 \sqrt{2\pi}}{2\sigma} \exp\left (\frac{{\sigma_0}^2}{2{\sigma}^2} \right) \left[ 1- \erf \left(\frac{\sigma_0}{\sqrt{2}\sigma}\right)\right].
    \label{eq:6}
\end{equation}
The ratio $G(t-\tau_0)/G(t=\tau_0)$ can be used to fit the linewidth of the green curve in Fig.~\ref{fig:mi1}(c). The experimental linewidth is 93.5 ns and this is fitted by choosing $\sigma$ = 19.7 ns in $G(t-\tau_0)/G(t=\tau_0)$. Now $\eta$ can be determined from the green  curve's height of 0.475. This calculation yields the value of $\eta = 59.8$~\%. A greater than 50~\% forward scattering efficiency $\eta$ is expected as a fiber coupler mounted on a half-inch polished surface is used to hold the diffuser in place at the entrance aperture of the IS. Thus, probe photons are more likely to be scattered in the forward direction than in the backward direction.
The theoretical model described by  Eq.~(\ref{eq:5}) is plotted as the orange curve in Fig.~\ref{fig:theory}(b), which is in excellent agreement with the experimental data.

\section{Conclusions}
We present an experimental scheme that can mitigate the adverse effect of scattering on quantum correlations in a quantum system. We use MI as the information metric to quantify quantum correlations in our system, which consists of a two-mode squeezed state with 7~dB intensity-difference squeezing, generated by the FWM process in atomic $^{85}$Rb vapor. Our mitigation scheme involves only an IS, and we demonstrate a 47.5~\% MI recovery from scattering after 32.7~ns information storage time, despite experiencing an enormous photon loss of greater than 85~\%.

\textcolor{black}{Since scattering can contribute to both decoherence and loss -- the two main disruptive random processes in quantum systems, our scheme can thus be found useful in many circumstances where disruptive random processes are unavoidable, paving the way for implementing delicate QISE protocols in real-life situations. For instance, in quantum communication systems, scattering can cause quantum signals to deviate from their intended path or degrade in fidelity. Mitigating scattering helps maintain the coherence and integrity of quantum signals over long-distance communication channels, such as in quantum key distribution~\cite{valivarthi2016quantum} or teleportation protocols~\cite{lu2022quantum}. In quantum computing platforms, where delicate qubits are manipulated, scattering can introduce errors by prematurely decohering qubits~\cite{krojanski2006reduced}. Therefore, mitigating scattering effects is essential for improving the stability and coherence times of qubits, thereby enhancing the overall computational performance of quantum algorithms~\cite{schlosshauer2019quantum}. In quantum sensing systems, scattering can obstruct precise measurement techniques that rely on quantum states~\cite{matsuzaki2011magnetic}. For example, in quantum metrology using interferometric methods, minimizing scattering effects ensures accurate signal detections, thereby reliably improving the sensitivity of quantum sensors~\cite{taylor2016quantum}.}



\section*{acknowledgments}
The experiment was performed under the supervision of T.L. in his quantum sensing lab at the University of Tennessee at Chattanooga. Z. J. is supported by the Herman F. Heep and Minnie Belle Heep Texas A$\&$M University Endowed Fund held/administered by the Texas A$\&$M Foundation. We would like to thank the Robert A. Welch Foundation, grant No. A-1261, A-1547, and A-1943), Air Force Office of Scientific Research (Award No. FA9550-20-1-0366) and the National Science Foundation (Grant No. PHY-2013771). This material is also based upon work supported by the U.S. Department of Energy, Office of Science, Office of Biological and Environmental Research under Award Number DE-SC-0023103 and \textcolor{black}{DE-AC36-08GO28308}.






\bibliography{QScattering}

\begin{thebibliography}{37}%
\makeatletter
\providecommand \@ifxundefined [1]{%
 \@ifx{#1\undefined}
}%
\providecommand \@ifnum [1]{%
 \ifnum #1\expandafter \@firstoftwo
 \else \expandafter \@secondoftwo
 \fi
}%
\providecommand \@ifx [1]{%
 \ifx #1\expandafter \@firstoftwo
 \else \expandafter \@secondoftwo
 \fi
}%
\providecommand \natexlab [1]{#1}%
\providecommand \enquote  [1]{``#1''}%
\providecommand \bibnamefont  [1]{#1}%
\providecommand \bibfnamefont [1]{#1}%
\providecommand \citenamefont [1]{#1}%
\providecommand \href@noop [0]{\@secondoftwo}%
\providecommand \href [0]{\begingroup \@sanitize@url \@href}%
\providecommand \@href[1]{\@@startlink{#1}\@@href}%
\providecommand \@@href[1]{\endgroup#1\@@endlink}%
\providecommand \@sanitize@url [0]{\catcode `\\12\catcode `\$12\catcode `\&12\catcode `\#12\catcode `\^12\catcode `\_12\catcode `\%12\relax}%
\providecommand \@@startlink[1]{}%
\providecommand \@@endlink[0]{}%
\providecommand \url  [0]{\begingroup\@sanitize@url \@url }%
\providecommand \@url [1]{\endgroup\@href {#1}{\urlprefix }}%
\providecommand \urlprefix  [0]{URL }%
\providecommand \Eprint [0]{\href }%
\providecommand \doibase [0]{https://doi.org/}%
\providecommand \selectlanguage [0]{\@gobble}%
\providecommand \bibinfo  [0]{\@secondoftwo}%
\providecommand \bibfield  [0]{\@secondoftwo}%
\providecommand \translation [1]{[#1]}%
\providecommand \BibitemOpen [0]{}%
\providecommand \bibitemStop [0]{}%
\providecommand \bibitemNoStop [0]{.\EOS\space}%
\providecommand \EOS [0]{\spacefactor3000\relax}%
\providecommand \BibitemShut  [1]{\csname bibitem#1\endcsname}%
\let\auto@bib@innerbib\@empty
\bibitem [{\citenamefont {Nielsen}\ and\ \citenamefont {Chuang}(2010)}]{nielsen2010quantum}%
  \BibitemOpen
  \bibfield  {author} {\bibinfo {author} {\bibfnamefont {M.~A.}\ \bibnamefont {Nielsen}}\ and\ \bibinfo {author} {\bibfnamefont {I.~L.}\ \bibnamefont {Chuang}},\ }\href@noop {} {\emph {\bibinfo {title} {Quantum computation and quantum information}}}\ (\bibinfo  {publisher} {Cambridge university press},\ \bibinfo {year} {2010})\BibitemShut {NoStop}%
\bibitem [{\citenamefont {Joos}\ \emph {et~al.}(2013)\citenamefont {Joos}, \citenamefont {Zeh}, \citenamefont {Kiefer}, \citenamefont {Giulini}, \citenamefont {Kupsch},\ and\ \citenamefont {Stamatescu}}]{joos2013decoherence}%
  \BibitemOpen
  \bibfield  {author} {\bibinfo {author} {\bibfnamefont {E.}~\bibnamefont {Joos}}, \bibinfo {author} {\bibfnamefont {H.~D.}\ \bibnamefont {Zeh}}, \bibinfo {author} {\bibfnamefont {C.}~\bibnamefont {Kiefer}}, \bibinfo {author} {\bibfnamefont {D.~J.}\ \bibnamefont {Giulini}}, \bibinfo {author} {\bibfnamefont {J.}~\bibnamefont {Kupsch}},\ and\ \bibinfo {author} {\bibfnamefont {I.-O.}\ \bibnamefont {Stamatescu}},\ }\href@noop {} {\emph {\bibinfo {title} {Decoherence and the appearance of a classical world in quantum theory}}}\ (\bibinfo  {publisher} {Springer Science \& Business Media},\ \bibinfo {year} {2013})\BibitemShut {NoStop}%
\bibitem [{\citenamefont {Bachor}\ and\ \citenamefont {Ralph}(2019)}]{bachor2019guide}%
  \BibitemOpen
  \bibfield  {author} {\bibinfo {author} {\bibfnamefont {H.-A.}\ \bibnamefont {Bachor}}\ and\ \bibinfo {author} {\bibfnamefont {T.~C.}\ \bibnamefont {Ralph}},\ }\href@noop {} {\emph {\bibinfo {title} {A guide to experiments in quantum optics}}}\ (\bibinfo  {publisher} {John Wiley \& Sons},\ \bibinfo {year} {2019})\BibitemShut {NoStop}%
\bibitem [{\citenamefont {Li}\ \emph {et~al.}(2022)\citenamefont {Li}, \citenamefont {Li}, \citenamefont {Liu}, \citenamefont {Yakovlev},\ and\ \citenamefont {Agarwal}}]{li2022quantum}%
  \BibitemOpen
  \bibfield  {author} {\bibinfo {author} {\bibfnamefont {T.}~\bibnamefont {Li}}, \bibinfo {author} {\bibfnamefont {F.}~\bibnamefont {Li}}, \bibinfo {author} {\bibfnamefont {X.}~\bibnamefont {Liu}}, \bibinfo {author} {\bibfnamefont {V.~V.}\ \bibnamefont {Yakovlev}},\ and\ \bibinfo {author} {\bibfnamefont {G.~S.}\ \bibnamefont {Agarwal}},\ }\bibfield  {title} {\bibinfo {title} {Quantum-enhanced stimulated brillouin scattering spectroscopy and imaging},\ }\href@noop {} {\bibfield  {journal} {\bibinfo  {journal} {Optica}\ }\textbf {\bibinfo {volume} {9}},\ \bibinfo {pages} {959} (\bibinfo {year} {2022})}\BibitemShut {NoStop}%
\bibitem [{\citenamefont {Casacio}\ \emph {et~al.}(2021)\citenamefont {Casacio}, \citenamefont {Madsen}, \citenamefont {Terrasson}, \citenamefont {Waleed}, \citenamefont {Barnscheidt}, \citenamefont {Hage}, \citenamefont {Taylor},\ and\ \citenamefont {Bowen}}]{casacio2021quantum}%
  \BibitemOpen
  \bibfield  {author} {\bibinfo {author} {\bibfnamefont {C.~A.}\ \bibnamefont {Casacio}}, \bibinfo {author} {\bibfnamefont {L.~S.}\ \bibnamefont {Madsen}}, \bibinfo {author} {\bibfnamefont {A.}~\bibnamefont {Terrasson}}, \bibinfo {author} {\bibfnamefont {M.}~\bibnamefont {Waleed}}, \bibinfo {author} {\bibfnamefont {K.}~\bibnamefont {Barnscheidt}}, \bibinfo {author} {\bibfnamefont {B.}~\bibnamefont {Hage}}, \bibinfo {author} {\bibfnamefont {M.~A.}\ \bibnamefont {Taylor}},\ and\ \bibinfo {author} {\bibfnamefont {W.~P.}\ \bibnamefont {Bowen}},\ }\bibfield  {title} {\bibinfo {title} {Quantum-enhanced nonlinear microscopy},\ }\href@noop {} {\bibfield  {journal} {\bibinfo  {journal} {Nature}\ }\textbf {\bibinfo {volume} {594}},\ \bibinfo {pages} {201} (\bibinfo {year} {2021})}\BibitemShut {NoStop}%
\bibitem [{\citenamefont {de~Andrade}\ \emph {et~al.}(2020)\citenamefont {de~Andrade}, \citenamefont {Kerdoncuff}, \citenamefont {Berg-S{\o}rensen}, \citenamefont {Gehring}, \citenamefont {Lassen},\ and\ \citenamefont {Andersen}}]{de2020quantum}%
  \BibitemOpen
  \bibfield  {author} {\bibinfo {author} {\bibfnamefont {R.~B.}\ \bibnamefont {de~Andrade}}, \bibinfo {author} {\bibfnamefont {H.}~\bibnamefont {Kerdoncuff}}, \bibinfo {author} {\bibfnamefont {K.}~\bibnamefont {Berg-S{\o}rensen}}, \bibinfo {author} {\bibfnamefont {T.}~\bibnamefont {Gehring}}, \bibinfo {author} {\bibfnamefont {M.}~\bibnamefont {Lassen}},\ and\ \bibinfo {author} {\bibfnamefont {U.~L.}\ \bibnamefont {Andersen}},\ }\bibfield  {title} {\bibinfo {title} {Quantum-enhanced continuous-wave stimulated raman scattering spectroscopy},\ }\href@noop {} {\bibfield  {journal} {\bibinfo  {journal} {Optica}\ }\textbf {\bibinfo {volume} {7}},\ \bibinfo {pages} {470} (\bibinfo {year} {2020})}\BibitemShut {NoStop}%
\bibitem [{\citenamefont {Lawrie}\ \emph {et~al.}(2019)\citenamefont {Lawrie}, \citenamefont {Lett}, \citenamefont {Marino},\ and\ \citenamefont {Pooser}}]{lawrie2019quantum}%
  \BibitemOpen
  \bibfield  {author} {\bibinfo {author} {\bibfnamefont {B.~J.}\ \bibnamefont {Lawrie}}, \bibinfo {author} {\bibfnamefont {P.~D.}\ \bibnamefont {Lett}}, \bibinfo {author} {\bibfnamefont {A.~M.}\ \bibnamefont {Marino}},\ and\ \bibinfo {author} {\bibfnamefont {R.~C.}\ \bibnamefont {Pooser}},\ }\bibfield  {title} {\bibinfo {title} {Quantum sensing with squeezed light},\ }\href@noop {} {\bibfield  {journal} {\bibinfo  {journal} {Acs Photonics}\ }\textbf {\bibinfo {volume} {6}},\ \bibinfo {pages} {1307} (\bibinfo {year} {2019})}\BibitemShut {NoStop}%
\bibitem [{\citenamefont {Michael}\ \emph {et~al.}(2019)\citenamefont {Michael}, \citenamefont {Bello}, \citenamefont {Rosenbluh},\ and\ \citenamefont {Pe’er}}]{michael2019squeezing}%
  \BibitemOpen
  \bibfield  {author} {\bibinfo {author} {\bibfnamefont {Y.}~\bibnamefont {Michael}}, \bibinfo {author} {\bibfnamefont {L.}~\bibnamefont {Bello}}, \bibinfo {author} {\bibfnamefont {M.}~\bibnamefont {Rosenbluh}},\ and\ \bibinfo {author} {\bibfnamefont {A.}~\bibnamefont {Pe’er}},\ }\bibfield  {title} {\bibinfo {title} {Squeezing-enhanced raman spectroscopy},\ }\href@noop {} {\bibfield  {journal} {\bibinfo  {journal} {npj Quantum Information}\ }\textbf {\bibinfo {volume} {5}},\ \bibinfo {pages} {81} (\bibinfo {year} {2019})}\BibitemShut {NoStop}%
\bibitem [{\citenamefont {Clark}\ \emph {et~al.}(2016)\citenamefont {Clark}, \citenamefont {Lecocq}, \citenamefont {Simmonds}, \citenamefont {Aumentado},\ and\ \citenamefont {Teufel}}]{clark2016observation}%
  \BibitemOpen
  \bibfield  {author} {\bibinfo {author} {\bibfnamefont {J.~B.}\ \bibnamefont {Clark}}, \bibinfo {author} {\bibfnamefont {F.}~\bibnamefont {Lecocq}}, \bibinfo {author} {\bibfnamefont {R.~W.}\ \bibnamefont {Simmonds}}, \bibinfo {author} {\bibfnamefont {J.}~\bibnamefont {Aumentado}},\ and\ \bibinfo {author} {\bibfnamefont {J.~D.}\ \bibnamefont {Teufel}},\ }\bibfield  {title} {\bibinfo {title} {Observation of strong radiation pressure forces from squeezed light on a mechanical oscillator},\ }\href@noop {} {\bibfield  {journal} {\bibinfo  {journal} {Nature Physics}\ }\textbf {\bibinfo {volume} {12}},\ \bibinfo {pages} {683} (\bibinfo {year} {2016})}\BibitemShut {NoStop}%
\bibitem [{\citenamefont {Aasi}\ \emph {et~al.}(2013)\citenamefont {Aasi}, \citenamefont {Abadie}, \citenamefont {Abbott}, \citenamefont {Abbott}, \citenamefont {Abbott}, \citenamefont {Abernathy}, \citenamefont {Adams}, \citenamefont {Adams}, \citenamefont {Addesso}, \citenamefont {Adhikari} \emph {et~al.}}]{aasi2013enhanced}%
  \BibitemOpen
  \bibfield  {author} {\bibinfo {author} {\bibfnamefont {J.}~\bibnamefont {Aasi}}, \bibinfo {author} {\bibfnamefont {J.}~\bibnamefont {Abadie}}, \bibinfo {author} {\bibfnamefont {B.}~\bibnamefont {Abbott}}, \bibinfo {author} {\bibfnamefont {R.}~\bibnamefont {Abbott}}, \bibinfo {author} {\bibfnamefont {T.}~\bibnamefont {Abbott}}, \bibinfo {author} {\bibfnamefont {M.}~\bibnamefont {Abernathy}}, \bibinfo {author} {\bibfnamefont {C.}~\bibnamefont {Adams}}, \bibinfo {author} {\bibfnamefont {T.}~\bibnamefont {Adams}}, \bibinfo {author} {\bibfnamefont {P.}~\bibnamefont {Addesso}}, \bibinfo {author} {\bibfnamefont {R.}~\bibnamefont {Adhikari}}, \emph {et~al.},\ }\bibfield  {title} {\bibinfo {title} {Enhanced sensitivity of the ligo gravitational wave detector by using squeezed states of light},\ }\href@noop {} {\bibfield  {journal} {\bibinfo  {journal} {Nature Photonics}\ }\textbf {\bibinfo {volume} {7}},\ \bibinfo {pages} {613} (\bibinfo {year} {2013})}\BibitemShut {NoStop}%
\bibitem [{\citenamefont {Taylor}\ \emph {et~al.}(2013)\citenamefont {Taylor}, \citenamefont {Janousek}, \citenamefont {Daria}, \citenamefont {Knittel}, \citenamefont {Hage}, \citenamefont {Bachor},\ and\ \citenamefont {Bowen}}]{taylor2013biological}%
  \BibitemOpen
  \bibfield  {author} {\bibinfo {author} {\bibfnamefont {M.~A.}\ \bibnamefont {Taylor}}, \bibinfo {author} {\bibfnamefont {J.}~\bibnamefont {Janousek}}, \bibinfo {author} {\bibfnamefont {V.}~\bibnamefont {Daria}}, \bibinfo {author} {\bibfnamefont {J.}~\bibnamefont {Knittel}}, \bibinfo {author} {\bibfnamefont {B.}~\bibnamefont {Hage}}, \bibinfo {author} {\bibfnamefont {H.-A.}\ \bibnamefont {Bachor}},\ and\ \bibinfo {author} {\bibfnamefont {W.~P.}\ \bibnamefont {Bowen}},\ }\bibfield  {title} {\bibinfo {title} {Biological measurement beyond the quantum limit},\ }\href@noop {} {\bibfield  {journal} {\bibinfo  {journal} {Nature Photonics}\ }\textbf {\bibinfo {volume} {7}},\ \bibinfo {pages} {229} (\bibinfo {year} {2013})}\BibitemShut {NoStop}%
\bibitem [{\citenamefont {Braunstein}\ and\ \citenamefont {Van~Loock}(2005)}]{braunstein2005quantum}%
  \BibitemOpen
  \bibfield  {author} {\bibinfo {author} {\bibfnamefont {S.~L.}\ \bibnamefont {Braunstein}}\ and\ \bibinfo {author} {\bibfnamefont {P.}~\bibnamefont {Van~Loock}},\ }\bibfield  {title} {\bibinfo {title} {Quantum information with continuous variables},\ }\href@noop {} {\bibfield  {journal} {\bibinfo  {journal} {Reviews of modern physics}\ }\textbf {\bibinfo {volume} {77}},\ \bibinfo {pages} {513} (\bibinfo {year} {2005})}\BibitemShut {NoStop}%
\bibitem [{\citenamefont {Wu}\ \emph {et~al.}(2019)\citenamefont {Wu}, \citenamefont {Liu}, \citenamefont {Chen},\ and\ \citenamefont {Chuu}}]{Wu2019}%
  \BibitemOpen
  \bibfield  {author} {\bibinfo {author} {\bibfnamefont {C.-H.}\ \bibnamefont {Wu}}, \bibinfo {author} {\bibfnamefont {C.-K.}\ \bibnamefont {Liu}}, \bibinfo {author} {\bibfnamefont {Y.-C.}\ \bibnamefont {Chen}},\ and\ \bibinfo {author} {\bibfnamefont {C.-S.}\ \bibnamefont {Chuu}},\ }\bibfield  {title} {\bibinfo {title} {Revival of quantum interference by modulating the biphotons},\ }\href {https://doi.org/10.1103/PhysRevLett.123.143601} {\bibfield  {journal} {\bibinfo  {journal} {Phys. Rev. Lett.}\ }\textbf {\bibinfo {volume} {123}},\ \bibinfo {pages} {143601} (\bibinfo {year} {2019})}\BibitemShut {NoStop}%
\bibitem [{\citenamefont {Serafini}\ \emph {et~al.}(2003)\citenamefont {Serafini}, \citenamefont {Illuminati},\ and\ \citenamefont {De~Siena}}]{serafini2003symplectic}%
  \BibitemOpen
  \bibfield  {author} {\bibinfo {author} {\bibfnamefont {A.}~\bibnamefont {Serafini}}, \bibinfo {author} {\bibfnamefont {F.}~\bibnamefont {Illuminati}},\ and\ \bibinfo {author} {\bibfnamefont {S.}~\bibnamefont {De~Siena}},\ }\bibfield  {title} {\bibinfo {title} {Symplectic invariants, entropic measures and correlations of gaussian states},\ }\href@noop {} {\bibfield  {journal} {\bibinfo  {journal} {Journal of Physics B: Atomic, Molecular and Optical Physics}\ }\textbf {\bibinfo {volume} {37}},\ \bibinfo {pages} {L21} (\bibinfo {year} {2003})}\BibitemShut {NoStop}%
\bibitem [{\citenamefont {Clark}\ \emph {et~al.}(2014)\citenamefont {Clark}, \citenamefont {Glasser}, \citenamefont {Glorieux}, \citenamefont {Vogl}, \citenamefont {Li}, \citenamefont {Jones},\ and\ \citenamefont {Lett}}]{clark2014quantum}%
  \BibitemOpen
  \bibfield  {author} {\bibinfo {author} {\bibfnamefont {J.~B.}\ \bibnamefont {Clark}}, \bibinfo {author} {\bibfnamefont {R.~T.}\ \bibnamefont {Glasser}}, \bibinfo {author} {\bibfnamefont {Q.}~\bibnamefont {Glorieux}}, \bibinfo {author} {\bibfnamefont {U.}~\bibnamefont {Vogl}}, \bibinfo {author} {\bibfnamefont {T.}~\bibnamefont {Li}}, \bibinfo {author} {\bibfnamefont {K.~M.}\ \bibnamefont {Jones}},\ and\ \bibinfo {author} {\bibfnamefont {P.~D.}\ \bibnamefont {Lett}},\ }\bibfield  {title} {\bibinfo {title} {Quantum mutual information of an entangled state propagating through a fast-light medium},\ }\href@noop {} {\bibfield  {journal} {\bibinfo  {journal} {Nature Photonics}\ }\textbf {\bibinfo {volume} {8}},\ \bibinfo {pages} {515} (\bibinfo {year} {2014})}\BibitemShut {NoStop}%
\bibitem [{\citenamefont {Vogl}\ \emph {et~al.}(2014)\citenamefont {Vogl}, \citenamefont {Glasser}, \citenamefont {Clark}, \citenamefont {Glorieux}, \citenamefont {Li}, \citenamefont {Corzo},\ and\ \citenamefont {Lett}}]{vogl2014advanced}%
  \BibitemOpen
  \bibfield  {author} {\bibinfo {author} {\bibfnamefont {U.}~\bibnamefont {Vogl}}, \bibinfo {author} {\bibfnamefont {R.~T.}\ \bibnamefont {Glasser}}, \bibinfo {author} {\bibfnamefont {J.~B.}\ \bibnamefont {Clark}}, \bibinfo {author} {\bibfnamefont {Q.}~\bibnamefont {Glorieux}}, \bibinfo {author} {\bibfnamefont {T.}~\bibnamefont {Li}}, \bibinfo {author} {\bibfnamefont {N.~V.}\ \bibnamefont {Corzo}},\ and\ \bibinfo {author} {\bibfnamefont {P.~D.}\ \bibnamefont {Lett}},\ }\bibfield  {title} {\bibinfo {title} {Advanced quantum noise correlations},\ }\href@noop {} {\bibfield  {journal} {\bibinfo  {journal} {New Journal of Physics}\ }\textbf {\bibinfo {volume} {16}},\ \bibinfo {pages} {013011} (\bibinfo {year} {2014})}\BibitemShut {NoStop}%
\bibitem [{\citenamefont {Weedbrook}\ \emph {et~al.}(2012)\citenamefont {Weedbrook}, \citenamefont {Pirandola}, \citenamefont {Garc{\'\i}a-Patr{\'o}n}, \citenamefont {Cerf}, \citenamefont {Ralph}, \citenamefont {Shapiro},\ and\ \citenamefont {Lloyd}}]{weedbrook2012gaussian}%
  \BibitemOpen
  \bibfield  {author} {\bibinfo {author} {\bibfnamefont {C.}~\bibnamefont {Weedbrook}}, \bibinfo {author} {\bibfnamefont {S.}~\bibnamefont {Pirandola}}, \bibinfo {author} {\bibfnamefont {R.}~\bibnamefont {Garc{\'\i}a-Patr{\'o}n}}, \bibinfo {author} {\bibfnamefont {N.~J.}\ \bibnamefont {Cerf}}, \bibinfo {author} {\bibfnamefont {T.~C.}\ \bibnamefont {Ralph}}, \bibinfo {author} {\bibfnamefont {J.~H.}\ \bibnamefont {Shapiro}},\ and\ \bibinfo {author} {\bibfnamefont {S.}~\bibnamefont {Lloyd}},\ }\bibfield  {title} {\bibinfo {title} {Gaussian quantum information},\ }\href@noop {} {\bibfield  {journal} {\bibinfo  {journal} {Reviews of Modern Physics}\ }\textbf {\bibinfo {volume} {84}},\ \bibinfo {pages} {621} (\bibinfo {year} {2012})}\BibitemShut {NoStop}%
\bibitem [{\citenamefont {Ameri}\ \emph {et~al.}(2015)\citenamefont {Ameri}, \citenamefont {Eghbali-Arani}, \citenamefont {Mari}, \citenamefont {Farace}, \citenamefont {Kheirandish}, \citenamefont {Giovannetti},\ and\ \citenamefont {Fazio}}]{ameri2015mutual}%
  \BibitemOpen
  \bibfield  {author} {\bibinfo {author} {\bibfnamefont {V.}~\bibnamefont {Ameri}}, \bibinfo {author} {\bibfnamefont {M.}~\bibnamefont {Eghbali-Arani}}, \bibinfo {author} {\bibfnamefont {A.}~\bibnamefont {Mari}}, \bibinfo {author} {\bibfnamefont {A.}~\bibnamefont {Farace}}, \bibinfo {author} {\bibfnamefont {F.}~\bibnamefont {Kheirandish}}, \bibinfo {author} {\bibfnamefont {V.}~\bibnamefont {Giovannetti}},\ and\ \bibinfo {author} {\bibfnamefont {R.}~\bibnamefont {Fazio}},\ }\bibfield  {title} {\bibinfo {title} {Mutual information as an order parameter for quantum synchronization},\ }\href@noop {} {\bibfield  {journal} {\bibinfo  {journal} {Physical Review A}\ }\textbf {\bibinfo {volume} {91}},\ \bibinfo {pages} {012301} (\bibinfo {year} {2015})}\BibitemShut {NoStop}%
\bibitem [{\citenamefont {Dixon}\ \emph {et~al.}(2012)\citenamefont {Dixon}, \citenamefont {Howland}, \citenamefont {Schneeloch},\ and\ \citenamefont {Howell}}]{dixon2012quantum}%
  \BibitemOpen
  \bibfield  {author} {\bibinfo {author} {\bibfnamefont {P.~B.}\ \bibnamefont {Dixon}}, \bibinfo {author} {\bibfnamefont {G.~A.}\ \bibnamefont {Howland}}, \bibinfo {author} {\bibfnamefont {J.}~\bibnamefont {Schneeloch}},\ and\ \bibinfo {author} {\bibfnamefont {J.~C.}\ \bibnamefont {Howell}},\ }\bibfield  {title} {\bibinfo {title} {Quantum mutual information capacity for high-dimensional entangled states},\ }\href@noop {} {\bibfield  {journal} {\bibinfo  {journal} {Physical review letters}\ }\textbf {\bibinfo {volume} {108}},\ \bibinfo {pages} {143603} (\bibinfo {year} {2012})}\BibitemShut {NoStop}%
\bibitem [{\citenamefont {Wolf}\ \emph {et~al.}(2008)\citenamefont {Wolf}, \citenamefont {Verstraete}, \citenamefont {Hastings},\ and\ \citenamefont {Cirac}}]{wolf2008area}%
  \BibitemOpen
  \bibfield  {author} {\bibinfo {author} {\bibfnamefont {M.~M.}\ \bibnamefont {Wolf}}, \bibinfo {author} {\bibfnamefont {F.}~\bibnamefont {Verstraete}}, \bibinfo {author} {\bibfnamefont {M.~B.}\ \bibnamefont {Hastings}},\ and\ \bibinfo {author} {\bibfnamefont {J.~I.}\ \bibnamefont {Cirac}},\ }\bibfield  {title} {\bibinfo {title} {Area laws in quantum systems: mutual information and correlations},\ }\href@noop {} {\bibfield  {journal} {\bibinfo  {journal} {Physical review letters}\ }\textbf {\bibinfo {volume} {100}},\ \bibinfo {pages} {070502} (\bibinfo {year} {2008})}\BibitemShut {NoStop}%
\bibitem [{\citenamefont {Belavkin}\ and\ \citenamefont {Ohya}(2002)}]{belavkin2002entanglement}%
  \BibitemOpen
  \bibfield  {author} {\bibinfo {author} {\bibfnamefont {V.~P.}\ \bibnamefont {Belavkin}}\ and\ \bibinfo {author} {\bibfnamefont {M.}~\bibnamefont {Ohya}},\ }\bibfield  {title} {\bibinfo {title} {Entanglement, quantum entropy and mutual information},\ }\href@noop {} {\bibfield  {journal} {\bibinfo  {journal} {Proceedings of the Royal Society of London. Series A: Mathematical, Physical and Engineering Sciences}\ }\textbf {\bibinfo {volume} {458}},\ \bibinfo {pages} {209} (\bibinfo {year} {2002})}\BibitemShut {NoStop}%
\bibitem [{\citenamefont {Dowran}\ \emph {et~al.}(2018)\citenamefont {Dowran}, \citenamefont {Kumar}, \citenamefont {Lawrie}, \citenamefont {Pooser},\ and\ \citenamefont {Marino}}]{dowran2018quantum}%
  \BibitemOpen
  \bibfield  {author} {\bibinfo {author} {\bibfnamefont {M.}~\bibnamefont {Dowran}}, \bibinfo {author} {\bibfnamefont {A.}~\bibnamefont {Kumar}}, \bibinfo {author} {\bibfnamefont {B.~J.}\ \bibnamefont {Lawrie}}, \bibinfo {author} {\bibfnamefont {R.~C.}\ \bibnamefont {Pooser}},\ and\ \bibinfo {author} {\bibfnamefont {A.~M.}\ \bibnamefont {Marino}},\ }\bibfield  {title} {\bibinfo {title} {Quantum-enhanced plasmonic sensing},\ }\href@noop {} {\bibfield  {journal} {\bibinfo  {journal} {Optica}\ }\textbf {\bibinfo {volume} {5}},\ \bibinfo {pages} {628} (\bibinfo {year} {2018})}\BibitemShut {NoStop}%
\bibitem [{\citenamefont {Anderson}\ \emph {et~al.}(2017)\citenamefont {Anderson}, \citenamefont {Gupta}, \citenamefont {Schmittberger}, \citenamefont {Horrom}, \citenamefont {Hermann-Avigliano}, \citenamefont {Jones},\ and\ \citenamefont {Lett}}]{anderson2017phase}%
  \BibitemOpen
  \bibfield  {author} {\bibinfo {author} {\bibfnamefont {B.~E.}\ \bibnamefont {Anderson}}, \bibinfo {author} {\bibfnamefont {P.}~\bibnamefont {Gupta}}, \bibinfo {author} {\bibfnamefont {B.~L.}\ \bibnamefont {Schmittberger}}, \bibinfo {author} {\bibfnamefont {T.}~\bibnamefont {Horrom}}, \bibinfo {author} {\bibfnamefont {C.}~\bibnamefont {Hermann-Avigliano}}, \bibinfo {author} {\bibfnamefont {K.~M.}\ \bibnamefont {Jones}},\ and\ \bibinfo {author} {\bibfnamefont {P.~D.}\ \bibnamefont {Lett}},\ }\bibfield  {title} {\bibinfo {title} {Phase sensing beyond the standard quantum limit with a variation on the su (1, 1) interferometer},\ }\href@noop {} {\bibfield  {journal} {\bibinfo  {journal} {Optica}\ }\textbf {\bibinfo {volume} {4}},\ \bibinfo {pages} {752} (\bibinfo {year} {2017})}\BibitemShut {NoStop}%
\bibitem [{\citenamefont {Pooser}\ and\ \citenamefont {Lawrie}(2015)}]{pooser2015ultrasensitive}%
  \BibitemOpen
  \bibfield  {author} {\bibinfo {author} {\bibfnamefont {R.~C.}\ \bibnamefont {Pooser}}\ and\ \bibinfo {author} {\bibfnamefont {B.}~\bibnamefont {Lawrie}},\ }\bibfield  {title} {\bibinfo {title} {Ultrasensitive measurement of microcantilever displacement below the shot-noise limit},\ }\href@noop {} {\bibfield  {journal} {\bibinfo  {journal} {Optica}\ }\textbf {\bibinfo {volume} {2}},\ \bibinfo {pages} {393} (\bibinfo {year} {2015})}\BibitemShut {NoStop}%
\bibitem [{\citenamefont {Hudelist}\ \emph {et~al.}(2014)\citenamefont {Hudelist}, \citenamefont {Kong}, \citenamefont {Liu}, \citenamefont {Jing}, \citenamefont {Ou},\ and\ \citenamefont {Zhang}}]{hudelist2014quantum}%
  \BibitemOpen
  \bibfield  {author} {\bibinfo {author} {\bibfnamefont {F.}~\bibnamefont {Hudelist}}, \bibinfo {author} {\bibfnamefont {J.}~\bibnamefont {Kong}}, \bibinfo {author} {\bibfnamefont {C.}~\bibnamefont {Liu}}, \bibinfo {author} {\bibfnamefont {J.}~\bibnamefont {Jing}}, \bibinfo {author} {\bibfnamefont {Z.}~\bibnamefont {Ou}},\ and\ \bibinfo {author} {\bibfnamefont {W.}~\bibnamefont {Zhang}},\ }\bibfield  {title} {\bibinfo {title} {Quantum metrology with parametric amplifier-based photon correlation interferometers},\ }\href@noop {} {\bibfield  {journal} {\bibinfo  {journal} {Nature communications}\ }\textbf {\bibinfo {volume} {5}},\ \bibinfo {pages} {3049} (\bibinfo {year} {2014})}\BibitemShut {NoStop}%
\bibitem [{\citenamefont {Li}\ \emph {et~al.}(2021)\citenamefont {Li}, \citenamefont {Li},\ and\ \citenamefont {Agarwal}}]{li2021experimental}%
  \BibitemOpen
  \bibfield  {author} {\bibinfo {author} {\bibfnamefont {F.}~\bibnamefont {Li}}, \bibinfo {author} {\bibfnamefont {T.}~\bibnamefont {Li}},\ and\ \bibinfo {author} {\bibfnamefont {G.~S.}\ \bibnamefont {Agarwal}},\ }\bibfield  {title} {\bibinfo {title} {Experimental study of decoherence of the two-mode squeezed vacuum state via second harmonic generation},\ }\href@noop {} {\bibfield  {journal} {\bibinfo  {journal} {Physical Review Research}\ }\textbf {\bibinfo {volume} {3}},\ \bibinfo {pages} {033095} (\bibinfo {year} {2021})}\BibitemShut {NoStop}%
\bibitem [{\citenamefont {Li}\ \emph {et~al.}(2017)\citenamefont {Li}, \citenamefont {Anderson}, \citenamefont {Horrom}, \citenamefont {Schmittberger}, \citenamefont {Jones},\ and\ \citenamefont {Lett}}]{li2017improved}%
  \BibitemOpen
  \bibfield  {author} {\bibinfo {author} {\bibfnamefont {T.}~\bibnamefont {Li}}, \bibinfo {author} {\bibfnamefont {B.~E.}\ \bibnamefont {Anderson}}, \bibinfo {author} {\bibfnamefont {T.}~\bibnamefont {Horrom}}, \bibinfo {author} {\bibfnamefont {B.~L.}\ \bibnamefont {Schmittberger}}, \bibinfo {author} {\bibfnamefont {K.~M.}\ \bibnamefont {Jones}},\ and\ \bibinfo {author} {\bibfnamefont {P.~D.}\ \bibnamefont {Lett}},\ }\bibfield  {title} {\bibinfo {title} {Improved measurement of two-mode quantum correlations using a phase-sensitive amplifier},\ }\href@noop {} {\bibfield  {journal} {\bibinfo  {journal} {Optics Express}\ }\textbf {\bibinfo {volume} {25}},\ \bibinfo {pages} {21301} (\bibinfo {year} {2017})}\BibitemShut {NoStop}%
\bibitem [{\citenamefont {Vogl}\ \emph {et~al.}(2013)\citenamefont {Vogl}, \citenamefont {Glasser}, \citenamefont {Glorieux}, \citenamefont {Clark}, \citenamefont {Corzo},\ and\ \citenamefont {Lett}}]{vogl2013experimental}%
  \BibitemOpen
  \bibfield  {author} {\bibinfo {author} {\bibfnamefont {U.}~\bibnamefont {Vogl}}, \bibinfo {author} {\bibfnamefont {R.~T.}\ \bibnamefont {Glasser}}, \bibinfo {author} {\bibfnamefont {Q.}~\bibnamefont {Glorieux}}, \bibinfo {author} {\bibfnamefont {J.~B.}\ \bibnamefont {Clark}}, \bibinfo {author} {\bibfnamefont {N.~V.}\ \bibnamefont {Corzo}},\ and\ \bibinfo {author} {\bibfnamefont {P.~D.}\ \bibnamefont {Lett}},\ }\bibfield  {title} {\bibinfo {title} {Experimental characterization of gaussian quantum discord generated by four-wave mixing},\ }\href@noop {} {\bibfield  {journal} {\bibinfo  {journal} {Physical Review A}\ }\textbf {\bibinfo {volume} {87}},\ \bibinfo {pages} {010101} (\bibinfo {year} {2013})}\BibitemShut {NoStop}%
\bibitem [{\citenamefont {Li}\ \emph {et~al.}(2020)\citenamefont {Li}, \citenamefont {Li}, \citenamefont {Altuzarra}, \citenamefont {Classen},\ and\ \citenamefont {Agarwal}}]{li2020squeezed}%
  \BibitemOpen
  \bibfield  {author} {\bibinfo {author} {\bibfnamefont {T.}~\bibnamefont {Li}}, \bibinfo {author} {\bibfnamefont {F.}~\bibnamefont {Li}}, \bibinfo {author} {\bibfnamefont {C.}~\bibnamefont {Altuzarra}}, \bibinfo {author} {\bibfnamefont {A.}~\bibnamefont {Classen}},\ and\ \bibinfo {author} {\bibfnamefont {G.~S.}\ \bibnamefont {Agarwal}},\ }\bibfield  {title} {\bibinfo {title} {Squeezed light induced two-photon absorption fluorescence of fluorescein biomarkers},\ }\href@noop {} {\bibfield  {journal} {\bibinfo  {journal} {Applied Physics Letters}\ }\textbf {\bibinfo {volume} {116}} (\bibinfo {year} {2020})}\BibitemShut {NoStop}%
\bibitem [{\citenamefont {Fry}\ \emph {et~al.}(2006)\citenamefont {Fry}, \citenamefont {Musser}, \citenamefont {Kattawar},\ and\ \citenamefont {Zhai}}]{fry2006integrating}%
  \BibitemOpen
  \bibfield  {author} {\bibinfo {author} {\bibfnamefont {E.~S.}\ \bibnamefont {Fry}}, \bibinfo {author} {\bibfnamefont {J.}~\bibnamefont {Musser}}, \bibinfo {author} {\bibfnamefont {G.~W.}\ \bibnamefont {Kattawar}},\ and\ \bibinfo {author} {\bibfnamefont {P.-W.}\ \bibnamefont {Zhai}},\ }\bibfield  {title} {\bibinfo {title} {Integrating cavities: temporal response},\ }\href@noop {} {\bibfield  {journal} {\bibinfo  {journal} {Applied optics}\ }\textbf {\bibinfo {volume} {45}},\ \bibinfo {pages} {9053} (\bibinfo {year} {2006})}\BibitemShut {NoStop}%
\bibitem [{\citenamefont {Duan}\ \emph {et~al.}(2000)\citenamefont {Duan}, \citenamefont {Giedke}, \citenamefont {Cirac},\ and\ \citenamefont {Zoller}}]{duan2000inseparability}%
  \BibitemOpen
  \bibfield  {author} {\bibinfo {author} {\bibfnamefont {L.-M.}\ \bibnamefont {Duan}}, \bibinfo {author} {\bibfnamefont {G.}~\bibnamefont {Giedke}}, \bibinfo {author} {\bibfnamefont {J.~I.}\ \bibnamefont {Cirac}},\ and\ \bibinfo {author} {\bibfnamefont {P.}~\bibnamefont {Zoller}},\ }\bibfield  {title} {\bibinfo {title} {Inseparability criterion for continuous variable systems},\ }\href@noop {} {\bibfield  {journal} {\bibinfo  {journal} {Physical review letters}\ }\textbf {\bibinfo {volume} {84}},\ \bibinfo {pages} {2722} (\bibinfo {year} {2000})}\BibitemShut {NoStop}%
\bibitem [{\citenamefont {Valivarthi}\ \emph {et~al.}(2016)\citenamefont {Valivarthi}, \citenamefont {Puigibert}, \citenamefont {Zhou}, \citenamefont {Aguilar}, \citenamefont {Verma}, \citenamefont {Marsili}, \citenamefont {Shaw}, \citenamefont {Nam}, \citenamefont {Oblak},\ and\ \citenamefont {Tittel}}]{valivarthi2016quantum}%
  \BibitemOpen
  \bibfield  {author} {\bibinfo {author} {\bibfnamefont {R.}~\bibnamefont {Valivarthi}}, \bibinfo {author} {\bibfnamefont {M.~l.~G.}\ \bibnamefont {Puigibert}}, \bibinfo {author} {\bibfnamefont {Q.}~\bibnamefont {Zhou}}, \bibinfo {author} {\bibfnamefont {G.~H.}\ \bibnamefont {Aguilar}}, \bibinfo {author} {\bibfnamefont {V.~B.}\ \bibnamefont {Verma}}, \bibinfo {author} {\bibfnamefont {F.}~\bibnamefont {Marsili}}, \bibinfo {author} {\bibfnamefont {M.~D.}\ \bibnamefont {Shaw}}, \bibinfo {author} {\bibfnamefont {S.~W.}\ \bibnamefont {Nam}}, \bibinfo {author} {\bibfnamefont {D.}~\bibnamefont {Oblak}},\ and\ \bibinfo {author} {\bibfnamefont {W.}~\bibnamefont {Tittel}},\ }\bibfield  {title} {\bibinfo {title} {Quantum teleportation across a metropolitan fibre network},\ }\href@noop {} {\bibfield  {journal} {\bibinfo  {journal} {Nature Photonics}\ }\textbf {\bibinfo {volume} {10}},\ \bibinfo {pages} {676} (\bibinfo {year} {2016})}\BibitemShut {NoStop}%
\bibitem [{\citenamefont {Lu}\ \emph {et~al.}(2022)\citenamefont {Lu}, \citenamefont {Wang}, \citenamefont {Huang}, \citenamefont {Wu}, \citenamefont {Wang}, \citenamefont {Wang}, \citenamefont {He}, \citenamefont {Yin}, \citenamefont {Guo}, \citenamefont {Chen} \emph {et~al.}}]{lu2022quantum}%
  \BibitemOpen
  \bibfield  {author} {\bibinfo {author} {\bibfnamefont {Q.-H.}\ \bibnamefont {Lu}}, \bibinfo {author} {\bibfnamefont {F.-X.}\ \bibnamefont {Wang}}, \bibinfo {author} {\bibfnamefont {K.}~\bibnamefont {Huang}}, \bibinfo {author} {\bibfnamefont {X.}~\bibnamefont {Wu}}, \bibinfo {author} {\bibfnamefont {Z.-H.}\ \bibnamefont {Wang}}, \bibinfo {author} {\bibfnamefont {S.}~\bibnamefont {Wang}}, \bibinfo {author} {\bibfnamefont {D.-Y.}\ \bibnamefont {He}}, \bibinfo {author} {\bibfnamefont {Z.-Q.}\ \bibnamefont {Yin}}, \bibinfo {author} {\bibfnamefont {G.-C.}\ \bibnamefont {Guo}}, \bibinfo {author} {\bibfnamefont {W.}~\bibnamefont {Chen}}, \emph {et~al.},\ }\bibfield  {title} {\bibinfo {title} {Quantum key distribution over a channel with scattering},\ }\href@noop {} {\bibfield  {journal} {\bibinfo  {journal} {Physical Review Applied}\ }\textbf {\bibinfo {volume} {17}},\ \bibinfo {pages} {034045} (\bibinfo {year} {2022})}\BibitemShut {NoStop}%
\bibitem [{\citenamefont {Krojanski}\ and\ \citenamefont {Suter}(2006)}]{krojanski2006reduced}%
  \BibitemOpen
  \bibfield  {author} {\bibinfo {author} {\bibfnamefont {H.~G.}\ \bibnamefont {Krojanski}}\ and\ \bibinfo {author} {\bibfnamefont {D.}~\bibnamefont {Suter}},\ }\bibfield  {title} {\bibinfo {title} {Reduced decoherence in large quantum registers},\ }\href@noop {} {\bibfield  {journal} {\bibinfo  {journal} {Physical review letters}\ }\textbf {\bibinfo {volume} {97}},\ \bibinfo {pages} {150503} (\bibinfo {year} {2006})}\BibitemShut {NoStop}%
\bibitem [{\citenamefont {Schlosshauer}(2019)}]{schlosshauer2019quantum}%
  \BibitemOpen
  \bibfield  {author} {\bibinfo {author} {\bibfnamefont {M.}~\bibnamefont {Schlosshauer}},\ }\bibfield  {title} {\bibinfo {title} {Quantum decoherence},\ }\href@noop {} {\bibfield  {journal} {\bibinfo  {journal} {Physics Reports}\ }\textbf {\bibinfo {volume} {831}},\ \bibinfo {pages} {1} (\bibinfo {year} {2019})}\BibitemShut {NoStop}%
\bibitem [{\citenamefont {Matsuzaki}\ \emph {et~al.}(2011)\citenamefont {Matsuzaki}, \citenamefont {Benjamin},\ and\ \citenamefont {Fitzsimons}}]{matsuzaki2011magnetic}%
  \BibitemOpen
  \bibfield  {author} {\bibinfo {author} {\bibfnamefont {Y.}~\bibnamefont {Matsuzaki}}, \bibinfo {author} {\bibfnamefont {S.~C.}\ \bibnamefont {Benjamin}},\ and\ \bibinfo {author} {\bibfnamefont {J.}~\bibnamefont {Fitzsimons}},\ }\bibfield  {title} {\bibinfo {title} {Magnetic field sensing beyond the standard quantum limit under the effect of decoherence},\ }\href@noop {} {\bibfield  {journal} {\bibinfo  {journal} {Physical Review A—Atomic, Molecular, and Optical Physics}\ }\textbf {\bibinfo {volume} {84}},\ \bibinfo {pages} {012103} (\bibinfo {year} {2011})}\BibitemShut {NoStop}%
\bibitem [{\citenamefont {Taylor}\ and\ \citenamefont {Bowen}(2016)}]{taylor2016quantum}%
  \BibitemOpen
  \bibfield  {author} {\bibinfo {author} {\bibfnamefont {M.~A.}\ \bibnamefont {Taylor}}\ and\ \bibinfo {author} {\bibfnamefont {W.~P.}\ \bibnamefont {Bowen}},\ }\bibfield  {title} {\bibinfo {title} {Quantum metrology and its application in biology},\ }\href@noop {} {\bibfield  {journal} {\bibinfo  {journal} {Physics Reports}\ }\textbf {\bibinfo {volume} {615}},\ \bibinfo {pages} {1} (\bibinfo {year} {2016})}\BibitemShut {NoStop}%
\end{thebibliography}%
\end{document}